\documentstyle[preprint,prb,aps]{revtex}
\begin{document}
\draft
\title{Comment on ``Theory of electron energy loss in a random
system of spheres''}
\author{Liang Fu and Lorenzo Resca}
\address{Department of Physics, the Catholic University of America,
Washington, D.C. 20064}
\maketitle
\begin{abstract}

In a recent paper, Barrera and Fuchs [Phys. Rev. B {\bf 52}, 3256
(1995)] provide a theory of electron energy loss in a random
system of spheres.  In this comment, we point out that the
approach of Barrera and Fuchs (A) completely ignores magnetic
excitations and retardation, which is inconsistent with the data
of energy loss of fast electrons, (B) does not yield the correct
$k\rightarrow 0$ limit, hence, does not recover correctly the
Maxwell-Garnett formula, and (C) does not use the correct screened
potential at any $k$.

\end{abstract}
\pacs{61.80.Mk, 77.22.-d, 77.84.Lf}

A major point of a recent paper by Barrera and Fuchs on the
electron energy loss in a composite medium with randomly
distributed spheres is to determine the effective dielectric
function of the composite with a mean-field theory that includes
all multipoles and the two-particle distribution.\cite{Barrera} 
This has already been done in the quasistatic case,
$k=0$.\cite{Paper1}  Barrera and Fuchs attempt a generalization of
that to finite $k$.  They retain the quasistatic multipolar
expansion, the corresponding re-expansion formula, and
the mean-field average procedure already developed,\cite{Paper1}
and simply replace the external potential $-E_0z$ with
$V_0e^{ikz}$.  Then, they claim that, in the limit $k\rightarrow
0$, they recover the Maxwell-Garnett formula (MGF), which was
demonstrated in the quasistatic case.\cite{Paper1} 

However, for the energy loss of fast electrons, with incident
energies of order 100 KeV,\cite{Barrera} the magnetic excitations
and the retardation are not negligible.  So, the treatment of Ref.
\onlinecite{Barrera} is more limited than a previous approach based
on the Poynting vector, in which magnetic excitations essentially
contribute to the energy loss.\cite{Pendry}  The complete $k\not=
0$ problem, including both electric and magnetic excitations, is
governed by Helmholtz equation: a full treatment of wave
propagation for periodic lattices of spheres has been provided in
Ref. \onlinecite{Pastori}, and in Ref. \onlinecite{McPhedran}
with electric excitations only.

Furthermore, the mean-field average of 
Ref. \onlinecite{Barrera} is incorrect,
and does not lead to the MGF for $k\rightarrow 0$.  We can show
that by demonstrating first what should be the correct mean-field
average and the correct derivation of 
the MGF for $k\rightarrow 0$. 

Starting from the central multipolar Eq. (16) of Ref.
\onlinecite{Barrera}, since we are only interested in the induced
dipole moment in the $k\rightarrow 0$ limit, we take $l=1$ and
expand $q_{10}^0$ for small $k$.  We then perform the mean-field
average, which yields 
\begin{eqnarray}
&&\langle q_{10i}\rangle =
-ikV_0\sqrt{3\over{4\pi}}\alpha_1\delta_l^1-
{{(2l+1)}\over{4\pi}}\alpha_1\sum_{l'}
\langle \sum_jB_{10i}^{l'0j}\rangle \langle q_{l'0j}\rangle .
\end{eqnarray}
These equations are basically those developed in Ref.
\onlinecite{Paper1}, with the replacement $E_0=-ikV_0$ for the
applied field (which is exactly the external field of Barrera and
Fuchs, for small $k$). Both the average coupling $\langle
\sum_jB_{10i}^{l'0j}\rangle$ and the average field depend on the
system configuration, and must be done  consistently to obtain the
correct effective dielectric function. In Ref.
\onlinecite{Paper1}, a parallel-plate configuration was introduced,
which correctly yields the MGF for isotropic  two-particle
distributions.  One of the advantages of the  parallel-plate
configuration is that the macroscopic  average field is simply the
potential difference between the electrode plates, divided by their
separation.  On the other hand, Ref. \onlinecite{Barrera}
effectively adopts a large spherical configuration, with a
hard-sphere two-particle distribution.  All the volume integrals
therein are indeed performed over such large sphere.  For such
system, the correct average must be
\begin{eqnarray}
&&\langle \sum_jB_{10i}^{l'0j}\rangle =\gamma_{1l'}\int\nolimits 
\rho^{(2)}(r)Y_{l'+1,0}(\theta )r^{-(l'+2)}d^3r\nonumber\\
&&=\gamma_{1l'}\int\nolimits_{0}^{\infty}{{\rho^{(2)}(r)}\over{r^
{l'}}}dr
\int\nolimits Y_{l'+1,0}(\theta)d\Omega =0, 
\end{eqnarray}
where $\gamma_{ll'}$ are given in Eq. (D8) of Ref.
\onlinecite{Barrera}.  Although the upper limit of the integral
over $r$ is taken to infinity, one must keep in mind that the
system, no matter how large, must be finite.  So, the proper
procedure must be that of doing the angular integrations first. 
Now, in the last term of Eq. (1), $-\sqrt{3/4\pi}\langle
\sum_jB_{10i}^{l'0j}\rangle \langle q_{l'0j}\rangle$ represents the
uniform part of the average field produced by the $l'$-order
multipoles of the surrounding particles onto the given central
particle.  That induces a dipole moment on the central particle, as
does the applied field, represented by the first term in the
right-hand side of Eq. (1).  So, our result (2) implies that a
spherical shell of uniformly distributed multipoles produces zero
field inside.  That is indeed the physically correct result,
consistent with the well known elementary example of a uniformly
charged spherical shell ($l'=0$), or that 
of a uniformly polarized
spherical shell ($l'=1$).  

Substituting Eq. (2) into Eq. (1), one obtains the average
polarization 
\begin{eqnarray}
\langle P\rangle = \sqrt{{4\pi}\over 3}n\langle q_{10}\rangle 
=-ikV_0n\tilde{\alpha}_1a^3,
\end{eqnarray}
where $\tilde{\alpha}_1=\alpha_1/a^3$.  Correspondingly, the
macroscopic average field for such a spherical system can be
obtained by solving the boundary-value problem of a homogeneous
sphere with a dielectric function $\epsilon_e$ in the applied
field $-ikV_0$, as done correctly in Eq. (8.3) of Ref.
\onlinecite{Felderhof}, for example, yielding 
\begin{eqnarray} \langle E\rangle
=-ikV_0{3\over{\epsilon_e+2}}. \end{eqnarray}  Then, from the
standard definition of the effective dielectric function (for a
vacuum host)
\begin{eqnarray}
\epsilon_e=1+4\pi \langle P\rangle /\langle E\rangle,
\end{eqnarray}
one obtains
\begin{eqnarray}
\epsilon_e={{1+2f\tilde{\alpha}_1}\over{1-f\tilde{\alpha}_1}},
\end{eqnarray}
where $f=4\pi na^3/3$ is the volume fraction of the particles. This
is exactly the MGF.  We emphasize again that in order to obtain
this result (MGF) correctly, both steps (2) and (4) must be done
consistently with the selected (spherical) system configuration. 

Now, according to the Eqs. (D11)-(D14) of Ref.
\onlinecite{Barrera}, 
the mean-field-average coupling
for the hard-sphere two-particle distribution
[$d\rho^{(2)}(r)/dr=n\delta
(r-2a)$] for arbitrary $k$ is
\begin{eqnarray}
\langle \sum_jB_{10i}^{l'0j}e^{ik(z_j-z_i)}\rangle =(-i)^{l'-1}
\langle \sum_j{\bar
B}_{10i}^{l'0j}e^{ik(z_j-z_i)}\rangle 
=(-i)^{l'-1}{{(4\pi )^2(l'+1)!}\over{
\sqrt{3(2l'+1)}l'!}}{{n}\over k}{{j_{l'}(k2a)}\over{(2a)^{l'}}}.
\end{eqnarray}
If the theory of Barrera and Fuchs were correct and 
reproduced the MGF in the $k\rightarrow 0$
limit, as they claim, Eq. (7) should vanish in such limit, as shown
in Eq. (2).  However, when the limit $k\rightarrow 0$ is taken, Eq.
(7) yields
instead 
\begin{eqnarray}
\langle \sum_jB_{10i}^{l'0j}\rangle '=(-i)^{l'-1}
\left\{{\begin{array}{ll} \infty ,\ \ &(l'=0)\\ 2n(4\pi /3)^2,\ \
&(l'=1)\\
0,\ \ &(l'>1).\end{array}}\right.
\end{eqnarray}
Here, we have used a prime to distinguish this incorrect result of
Ref. \onlinecite{Barrera} from our correct Eq. (2). The incorrect
result (8) would imply that a uniformly charged spherical shell,
with a {\it finite} and {\it real} charge density, would produces
inside 
an {\it infinite} and {\it imaginary} field, which is obviously
wrong. 
Likewise, a uniformly
polarized spherical shell would produce inside a field in the {\it
opposite} direction to the polarization, which is also clearly
incorrect. Furthermore, if the shape of the system is not
spherical, or the two-particle distribution is not spherical, the
higher multipoles ($l'>1$) of the surrounding particles will
contribute.\cite{Paper1} In that case, the incorrect $\langle
\sum_jB_{10i}^{l'0j}\rangle'$ would be imaginary or complex for
some $l'$.  Then, real higher multipoles would also produce 
complex fields on the central particle, which is definitely
incorrect.

There is another basic and independent error in the derivations 
of Ref. \onlinecite{Barrera}: namely, no distinction is made
between the fields screened and unscreened by the spherical system. 
This in itself is
enough to prevent a correct result for both finite and vanishing
$k$. 
Combining that
with the error in obtaining the mean-field averages, one cannot
expect to recover correctly the MGF.  Indeed, using the incorrect
result (8), one obtains for the average
polarization 
\begin{eqnarray}
\langle P\rangle'
=-ikV_0n{{\tilde{\alpha}_1a^3}\over{1+2f\tilde{\alpha}_1}}.
\end{eqnarray}
Then, it is easily verified that the incorrect
result (8) leads to
\begin{eqnarray}
\epsilon_e'={{1+4f\tilde{\alpha}_1}\over{1+f\tilde{\alpha}_1}},
\end{eqnarray}
if the screened field (4) is used, and to
\begin{eqnarray}
\epsilon_e'={{1+5f\tilde{\alpha}_1}\over{1+2f\tilde{\alpha}_1}},
\end{eqnarray}
if the unscreened field is used.  Evidently, neither of these
results (10) and (11) represents the correct MGF.  Nonetheless,
Barrera and Fuchs claim that they obtain that.

It may be worth noticing that, although Eq. (9) of Ref.
\onlinecite{Barrera} defines the effective dielectric function
through potentials, rather than electric fields, that has no effect
on the mean-field averages.  Nor does it avoid the screening
requirement on the potential, otherwise there would be no
justification to use any particular geometry, and 
the result would be completely arbitrary.

In summary, we have shown that the mean-field average of Ref.
\onlinecite{Barrera} is incorrect, does not reproduce the
Maxwell-Garnett formula in the $k\rightarrow 0$ limit, and does not
use the correct screened potential.  Combined with the limitations
of ignoring magnetic excitations and retardation in the response,
that leads to the conclusion that the relatively good fitting of
experimental data obtained in Ref. \onlinecite{Barrera} is at best
fortuitous, and simply reflects an arbitrary adjustment of
parameters, such as the volume fractions and the particle radii. 

This work has been supported in part by the U.S. Army Research
Office under contract No. DAAH04-93-G-0236.

\end{document}